\documentclass[aps,prb,twocolumn,showpacs,superscriptaddress,amsmath,amssymb,floatfix]{revtex4}

\usepackage{graphicx} 
\usepackage[a4paper,left=1.5cm,right=1cm,top=3cm,bottom=2cm]{geometry}
\usepackage{color}
\usepackage[colorlinks = false, citecolor=green, linkcolor=red]{hyperref}
\usepackage{dsfont}

\begin{document}
\title{Measuring central-spin interaction with a spin bath by pulsed ENDOR: \\ Towards suppression of spin diffusion decoherence}
\author{S.~J.~Balian}
\email[]{s.balian@ucl.ac.uk}
\affiliation{Department of Physics and Astronomy, University College London, Gower Street, London WC1E 6BT, United Kingdom}
\author{M.~B.~A.~Kunze}
\affiliation{Institute of Structural and Molecular Biology, University College London, Gower Street, London WC1E 6BT, United Kingdom}
\author{M.~H.~Mohammady}
\affiliation{Department of Physics and Astronomy, University College London, Gower Street, London WC1E 6BT, United Kingdom} 
\author{G.~W.~Morley}
 \affiliation{Department of Physics, University of Warwick, Gibbet Hill Road, Coventry CV4 7AL, United Kingdom}
\author{W.~M.~Witzel}
\affiliation{Sandia National Laboratories, Albuquerque, New Mexico 87185, USA}
\author{C.~W.~M.~Kay}
\affiliation{Institute of Structural and Molecular Biology, University College London, Gower Street, London WC1E 6BT, United Kingdom}
\affiliation{London Centre for Nanotechnology, University College London, 17-19 Gordon Street, London WC1H 0AH, United Kingdom}
\author{T.~S.~Monteiro}
\affiliation{Department of Physics and Astronomy, University College London, Gower Street, London WC1E 6BT, United Kingdom}
\date{\today}
\begin{abstract}

We present pulsed electron-nuclear double resonance (ENDOR) experiments which
enable us to characterize the coupling between bismuth donor spin qubits in Si
and the surrounding spin bath of $^{29}\text{Si}$ impurities which provides the 
dominant decoherence mechanism (nuclear spin diffusion) at low temperatures
($< 16$~K). Decoupling from the spin bath is predicted and cluster correlation
expansion simulations show near-complete suppression of spin diffusion, at
optimal working points. The suppression takes the form of sharply peaked
divergences of the spin diffusion coherence time, in contrast with previously
identified broader regions of insensitivity to classical fluctuations.
ENDOR data suggest that anisotropic contributions are comparatively weak,
so the form of the divergences is largely independent of crystal orientation.

\end{abstract}
\maketitle

\section{Introduction}\label{Sec:Intro}
Quantum decoherence presents a fundamental limitation to the realization of 
practical quantum computing and of other technological devices which actively
exploit quantum phenomena. In 2002, a ground-breaking study established the
usefulness of so-called optimal working points (OWPs):\cite{Vion2002} These
are parameter regimes where the system becomes - to first order - insensitive
to fluctuations of external classical fields. We consider here the effect of
OWPs in a system where decoherence of a central spin system arises from
interactions with a fluctuating bath of surrounding spins - a scenario that is
of considerable significance in the field of quantum information.
\cite{DeSousa2003_1, DeSousa2003_2, Abe2004, Takahashi2008, Abe2010, Witzel2010, BarGill2012, Zhao2012, DeLange2012}

A promising approach for silicon-based quantum-information processing (QIP)
involves combined electron and nuclear spins of donor atoms in Si, which are
amenable to high-fidelity manipulation by means of electron paramagnetic
resonance (EPR) and nuclear magnetic resonance (NMR), respectively.
Most studies have considered phosphorus ($^{31}\text{P}$) donors in Si.
\cite{Kane1998,Schofield2003,Stoneham2003,Tyryshkin2003,Fu2004,Morley2008,Morello2010,Greenland2010,Simmons2011,Steger2011,Dreher2012,Fuechsle2012}
More recently, several different groups have investigated another
Group V donor, $^{209}\text{Bi}$. These studies not only showed that bismuth
donors have similar properties to Si:P, such as long electron spin coherence times $T_2$ of
the order of several ms at low temperatures ($< 16$~K),\cite{Morley2010,George2010}
but that they also offer new possibilities for QIP. For example, strong optical
hyperpolarization was demonstrated,\cite{Morley2010,Sekiguchi2010} allowing
for efficient initialization of the nuclear spin.
The Si:Bi spin system has an electron spin $S=1/2$ and a large nuclear
spin $I=9/2$ as well as an atypically strong hyperfine coupling constant,
$A/2\pi=1.4754$~GHz. The strong state-mixing occurring for magnetic fields
$B\simeq 0.1-0.6$~T where the hyperfine interaction competes with the
electronic Zeeman energy allows transitions which are forbidden at high
magnetic fields,\cite{Mohammady2010,Mohammady2012} observed recently in
Ref.~\onlinecite{Morley2011}.
In Refs.~\onlinecite{Mohammady2010,Mohammady2012}, a set of minima and maxima
were found in the $f-B$ parameter space of dipole-allowed transitions at
frequencies $f$. These $df/dB\simeq0$ points were identified as OWPs: Line narrowing
and reduced sensitivity to temporal and spatial noise in $B$ over a broad
region of fields (closely related to $df/dB=0$ extrema) were found. However,
to date, their effectiveness for reducing decoherence in the real environment
of a spin bath remains untested.

In natural Si, 4.67\% of lattice sites are occupied by the nuclear spin-half
$^{29}\text{Si}$ isotope, rather than the spinless $^{28}\text{Si}$.
Flip-flopping of the $^{29}\text{Si}$ spins provides the dominant 
mechanism of decoherence for both Si:P and Si:Bi systems at low temperatures.
The decay of the donor Hahn spin echo for these systems is typically
fitted to $\text{exp}[-t/T_{2} - (t/T_{\text{SD}})^n ]$, where
$T_{\text{SD}} < 1$~ms characterizes the nuclear spin diffusion, with
$n \simeq 2 - 3$.\cite{Witzel2006} Other relaxation processes,
such as those arising from donor-donor interactions, are represented by $T_2$.
Since $T_2 \gg T_{\text{SD}}$, nuclear spin diffusion remains the main channel
of decoherence at low temperatures.\cite{Tyryshkin2006,Tyryshkin2012}

In this work, we investigate the nature of the Bi-$^{29}\text{Si}$ interaction
by means of pulsed electron-nuclear double resonance (ENDOR).\cite{Schweiger2001}
To obtain an ENDOR spectrum, an EPR spin echo is detected as a function of a
radio frequency (rf) excitation. When the rf radiation is resonant with an NMR
transition, changes are seen in the EPR signal if the populations of the
relevant energy levels change. Previous ENDOR studies of Si:Bi used rf
frequencies of at least several hundreds of MHz,\cite{Morley2010,George2010}
and thus could not probe the weak couplings to a spin-bath. In contrast, here,
rf frequencies of a few MHz were used. With this approach, we have measured
the bismuth spin bath superhyperfine (SHF) couplings and determined their
anisotropy.

We also present the results of cluster correlation expansion (CCE) simulations.
\cite{Witzel2006,Yang2008_2009} This model has been used with considerable
success to model central spin decoherence in Si:P.\cite{Abe2004,Abe2010,Witzel2010,Witzel2006}
In Ref.~\onlinecite{George2010}, weak state-mixing in Si:Bi was investigated by
simply allowing for the variation of an effective gyromagnetic ratio.
Here we adapt the CCE simulations to include, for the first time, the strong
state-mixing seen near the OWPs. A key finding is the demonstration of
near-complete suppression of nuclear spin diffusion, even in natural Si:
This occurs in extremely narrow regions, where $T_{\text{SD}}$ is in effect
divergent,\cite{Divergence} in contrast to the broader effect expected from the
form of $df/dB$.\cite{Vion2002} A successful means of controlling decoherence
is to employ isotopically enriched samples,\cite{Abe2004,Abe2010,Tyryshkin2003,Tyryshkin2006,Steger2011,Tyryshkin2012,Simmons2011,Steger2012,Weis2012,Wolfowicz2012}
which can exhibit long $T_2$ times up to the order of seconds.\cite{Tyryshkin2012}
Thus, the OWPs represent a potentially complementary technique, effective for both
natural Si and partially enriched samples. In addition, our work suggests that
the OWPs may also be effective in suppressing residual effects such as
donor-donor interactions which are responsible for shortening $T_2$ times
from seconds to the ms timescale.\cite{Mohammady2010}

\section{Spin Hamiltonian}\label{Sec:Hamiltonian}
We investigate a spin system with total Hamiltonian
\begin{equation}
\hat{H} =
A \hat{{\bf I}} \cdot \hat{{\bf S}}
+ \hat{H}_{\text{Zee}}
+ \hat{H}_{\text{int}}
+ \hat{H}_{\text{bath}}.
\label{Eq:Hamiltonian}
\end{equation}
The first term denotes the isotropic hyperfine interaction between the bismuth electronic
and nuclear spins.
For Si:Bi, the usual high-field reduction to Ising form 
$A \hat{{\bf I}} \cdot \hat{{\bf S}} \simeq A \hat{I}^z \hat{S}^z$
cannot be made at the fields of interest here.
The second term represents the Zeeman interaction with the external field,
including the donor spins and summed over bath spins,
\begin{equation}
\hat{H}_{\text{Zee}} = \omega_0 \left(\hat{S}^z - \delta_{\text{Bi}} \hat{I}^{z} - \delta_{\text{Si}} \sum_n \hat{I}_{n}^{z}\right).
\label{Eq:ZeemanH}
\end{equation}
Here, $\omega_0 = \mu B / \hbar$ is the electronic Zeeman frequency
($\mu  = 1.857 \times 10^{-23}$~JT$^{-1}$)
while $\delta_{\text{Bi}}=2.486\times 10^{-4}$
and $\delta_{\text{Si}}=3.021\times 10^{-4}$ are
the ratios of the nuclear to electronic Zeeman frequencies for the donor 
and $^{29}\text{Si}$ spins, respectively.

The spin-bath interaction term
\begin{equation}
\hat{H}_{\text{int}} = \sum_n \hat{{\bf I}}_{n} {\bf J}_n \hat{{\bf S}},
\label{Eq:InteractionH}
\end{equation}
represents the SHF couplings between the donor and bath spins, in general of
tensor form (for anisotropic couplings). 
Finally, dipolar coupling between each pair of $^{29}\text{Si}$ spins is
represented by the bath term
\begin{equation}
\hat{H}_{\text{bath}}=\sum_{n < m} \hat{{\bf I}}_{n} {\bf D}({\bf r}_{nm}) \hat{{\bf I}}_{m},
\label{Eq:BathH}
\end{equation}
where ${\bf r}_{nm}$ denotes the relative position vector of bath spins at
lattice sites $n$ and $m$. Writing ${\bf r}_{nm} \equiv {\bf r}$ for a pair of
spins, the components of the dipolar tensor ${\bf D}$ are given by
\begin{equation}
D_{i j}({\bf r}) = \left(\frac{\mu_0\delta_{\text{Si}}^2\mu^2}{4\pi\hbar r^3}\right)\left(\delta_{i j} - 3 \frac{r_{i} r_{j}}{r^2}\right),
\label{Eq:Dipolar}
\end{equation}
with $\mu_0 = 4\pi \times 10^{-7}$~NA$^{-2}$ and $i,j = x,y,z$.

\section{ENDOR measurements}\label{Sec:Experiment}
The experimental ENDOR studies reported here served to investigate and
characterize the isotropic/anisotropic character of the spin-bath interaction
term, namely a set of distinct ${\bf J}_n$ values - SHF couplings - in
Eq.~(\ref{Eq:InteractionH}), corresponding to occupancy of inequivalent lattice
sites by $^{29}\text{Si}$ impurities. Pulsed ENDOR experiments were performed
using the Davies ENDOR pulse sequence.\cite{Schweiger2001,Davies}
We applied the pulse sequence
$\pi_{\text{mw}}-\tau_1-\pi_{\text{rf}}-\tau_2-\frac{\pi}{2}{\text{\scriptsize mw}}-\tau_3-\pi_{\text{mw}}-\tau_3-\text{echo}$,
where the microwave (mw) frequency is chosen to excite one EPR
transition and the rf is stochastically varied between
$2-12$~MHz or $2-7$~MHz to excite all nuclear spin transitions
in this region. We used $256$~ns long $\pi_{\text{mw}}$-pulses and a $128$~ns
long $\frac{\pi}{2}{\text{\scriptsize mw}}$-pulse. For optimal signal-to-noise
ratio and resolution, we used a $\pi_{\text{rf}}$-pulse of $10~\mu\text{s}$.
Pulse delays were set to $\tau_1=1~\mu\text{s}$, $\tau_2=3~\mu\text{s}$, and
$\tau_3=1.5~\mu\text{s}$ and a shot repetition time of 1.3~ms was
employed to give a good signal-to-noise ratio. All experiments were carried out
at 15~K on an E580 pulsed EPR spectrometer (Bruker Biospin) equipped with
pulsed ENDOR accessory (E560D-P), a dielectric ring ENDOR resonator
(EN4118X-MD4), a liquid helium flow cryostat (Oxford CF935), and an rf amplifier
(ENI A-500W). We used a donor concentration of $3 \times 10^{15}$~cm$^{-3}$ and
the magnetic field was directed perpendicular to the [111] plane.

While not offering the higher frequency resolution attainable with
continuous-wave ENDOR,\cite{Feher1959,Hale1969} the pulsed ENDOR measurements
permit us to adequately constrain and to demonstrate the reliability of the
numerical simulations. In particular, we established that isotropic couplings
to the spin bath dominate the decoherence dynamics. While not the focus of this
study, a further motivation is to investigate the feasibility of an alternative
possibility for QIP: to simultaneously manipulate the $^{29}\text{Si}$ atoms as
spin-half qubits, along with the donors.\cite{Akhtar2012}

\begin{figure}[t]
\includegraphics[width=3.375in]{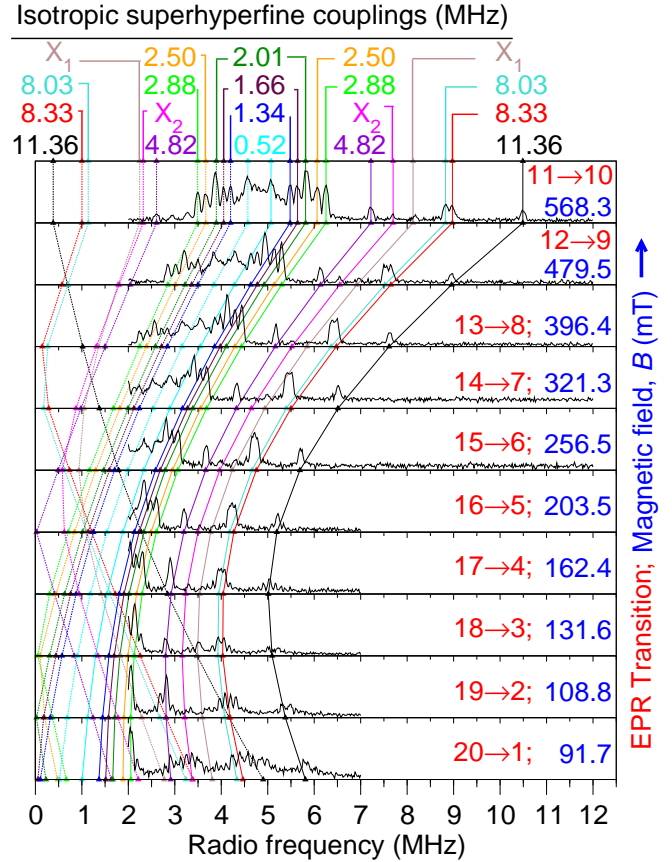}
\caption{(Color online) Pulsed ENDOR measured for bismuth-doped silicon with
frequency 9.8~GHz at which ten EPR lines are observed,
the resonance peaks due to interactions of the donor with $^{29}\text{Si}$
nuclei at inequivalent lattice sites. The isotropic superhyperfine couplings
were extracted from the spectrum at the highest magnetic field.
As the field is varied, the smooth lines follow the resonance positions
according to Eq.~(\ref{Eq:IsoENDOR}). Solid and dotted lines distinguish
between the two peaks observed for each coupling, each corresponding to one of
the two donor levels involved in the EPR transition.
Only the peaks labeled $X_1$ and $X_2$, in addition to a third pair
not resolved here, were found to show anisotropy from performing ENDOR as a
function of crystal orientation.}
\label{Fig:ENDOR}
\end{figure}

Measured ENDOR spectra at $f\simeq9.755$~GHz are presented in Fig.~\ref{Fig:ENDOR},
together with a list of SHF couplings. For the magnetic field range
$B \simeq 0.1-0.6$~T in Fig.~\ref{Fig:ENDOR}, there is significant mixing of
the high-field Si:Bi energy eigenstates $|m_{S}, m_{I}\rangle$. The mixed
eigenstates, $| \pm, m \rangle $, correspond to doublets (at most) of constant
$m=m_{S}+ m_{I}$:
\begin{eqnarray}
\left| \pm, m \right\rangle = a_{m}^{\pm}\left|\pm \tfrac{1}{2},m \mp \tfrac{1}{2} \right\rangle + b_{m}^{\pm} \left|\mp \tfrac{1}{2}, m \pm \tfrac{1}{2} \right\rangle, \\
 |a_{m}^{\pm}|^{2} - |b_{m}^{\pm}|^{2} = \frac{\Omega_{m}(\omega_0)}{\sqrt{\left(\Omega_{m}^{2}(\omega_0) + 25 - m^{2}\right)}} \equiv \gamma_{m}(\omega_0),
\label{Eq:SiBiStates}
\end{eqnarray}
where $\Omega_{m}(\omega_0)= m + \frac{\omega_0}{A}(1 + \delta_\text{Bi})$ and $m$ is an
integer, $-5\leq m \leq 5$. Such mixing leads to a complex EPR spectrum for
bismuth with $df/dB=0$ extrema. The minima correspond to transitions between
states corresponding to adjacent avoided level crossings, of which there are
four. The disparity between the electronic Zeeman and hyperfine energy scales
and SHF energy scales means that the tensor coupling in
Eq.~(\ref{Eq:InteractionH}) reduces to simpler form
\begin{equation}
\hat{H}_{\text{int},l} \approx ( \alpha_l \hat{I}^{z}_l + \beta_l \hat{I}^{x}_l ) \hat{S}^z,
\label{Eq:SimplifiedH}
\end{equation}
written for coupling to a single $^{29}\text{Si}$ at site $l$,
where $\alpha_l = [(a_{\text{iso},l}-T_l)+3T_l^{2}\cos^{2}\vartheta_l]$ and
$\beta_l = 3T_l\sin\vartheta_l\cos\vartheta_l$ with $a_{\text{iso},l}$ and
$T_l$ the isotropic and anisotropic parts of the molecular-frame SHF tensor,
respectively, and $\vartheta_l$ the angle between the external field and the
line connecting the bismuth site and site $l$. 
Nonsecular terms involving $\hat{S}^x$ and $\hat{S}^y$ can be neglected.\cite{Schweiger2001}
Diagonalization of the resulting 2-dimensional Hamiltonian, 
and setting $T_l=0$ for a purely isotropic coupling leads to a simple
expression for the ENDOR resonance frequency for donor level $|\pm,m\rangle$:
\begin{equation}
\Delta_{\text{iso},l}^{\pm,m}(\omega_0) = \frac{1}{2\pi}\left| - \omega_0  \delta_{\text{Si}} \pm \left(\frac{a_{\text{iso},l}}{2}\right) \gamma_{m}(\omega_0)\right|.
\label{Eq:IsoENDOR}
\end{equation}
The above expression is in perfect agreement with full numerical
diagonalization. The couplings in Fig.~\ref{Fig:ENDOR} were extracted from the
measured spectra by fitting to the data Gaussians of equal width and using
Eq.~(\ref{Eq:IsoENDOR}).
The same expression and a single set of couplings gave excellent agreement with
data at 10 different fields. In particular, the observed pattern
of half a dozen highest frequency $^{29}\text{Si}$ resonances moving to
a minimum at $B \simeq 0.2$~T, then increasing again, is directly attributable
to mixing of the states of the bismuth donor: i.e., here $\gamma_{m}(\omega_0)$
has a minimum.

Ten out of the twelve couplings extracted from data were found to be purely
isotropic. The highest-field spectrum was measured for a range of crystal
orientations and only three weak intensity lines showed orientation-dependent
frequencies and hence anisotropy. Two are indicated by $X_1$ and $X_2$ in
Fig.~\ref{Fig:ENDOR}: The corresponding two couplings with nonzero anisotropy
were found to have ($a_{\text{iso,$X_1$}}\simeq2.8$,~$T_{X_1}\simeq2.4$)~MHz and
($a_{\text{iso,$X_2$}}\simeq0.4$,~$T_{X_2}\simeq2.8$)~MHz by fitting the more general form of
Eq.~(\ref{Eq:IsoENDOR}) with nonzero $T$. A previous ESEEM (electron spin echo
envelope modulation) study identified a single anisotropic
coupling,\cite{Belli2011} attributed to $E$-shell (nearest neighbor)
$^{29}\text{Si}$. The third line we identify is fitted by coupling constants
consistent with the anisotropic coupling in Ref.~\onlinecite{Belli2011}.
For most crystal orientations, this line is masked by much higher intensity
lines arising from isotropic couplings.

At fields where $\gamma_{m}(\omega_0)$ becomes small (this occurs close to
the $df/dB=0$ minima as shown
in Refs.~\onlinecite{Mohammady2010, Mohammady2012}), Eq.~(\ref{Eq:IsoENDOR})
tends to the $^{29}\text{Si}$ Zeeman frequency $\delta_\text{Si}\omega_{0}$.
It is straightforward to extend Eq.~(\ref{Eq:IsoENDOR}) to the anisotropic case
and show that the latter statement also holds for anisotropic couplings.
In effect, at these points, the donor might be said to approximately decouple
from the bath. For example, for the EPR transition $|12\rangle \to |9\rangle$
(labeling the eigenstates $|n=1,2,\dots20\rangle$ in increasing order of
energy), $\gamma_{m}(\omega_0)=0$ at $B=157.9$~mT for level
$|12\rangle$ and $B=210.5$~mT for $|9\rangle$.
We note that there is however no $B$-field value where both the upper and lower
levels have $\gamma_{m}(\omega_0)=0$: As we see below, this is not 
actually essential for complete suppression of spin diffusion.
The actual OWP is at $B=188.0$~mT, where $\gamma_{-3}(\omega_0)=-\gamma_{-4}(\omega_0)$.
This is extremely close to where $df/dB=0$, which occurs when
\begin{equation}
\gamma_{-3}(\omega_0) + \gamma_{-4}(\omega_0) - \frac{2\delta_{\text{Bi}}}{(1+\delta_{\text{Bi}})} = 0.
\label{Eq:MinimumCondition}
\end{equation}
\section{Cluster correlation expansion simulations}\label{Sec:CCE}
In order to model the full dynamics, we assume that the temperature
and donor concentrations are low enough so that phonon-induced decoherence and
decoherence due to interactions between donors are negligible.
The Hahn spin echo decay of a central donor electron coupled to the bismuth
nucleus in a bath of $^{29}$Si was calculated using
the CCE.\cite{Yang2008_2009} Denoting the spin echo intensity by $L(t)$,
let $L_{\mathcal{S}}(t)$ be $L(t)$ computed including only spins in some set
or ``cluster'' of bath spins $\mathcal{S}$. The quantity
$\tilde{L}_{\mathcal{S}}(t)$ is defined as
\begin{equation}
L(t) = \prod_{\mathcal{S}} \tilde{L}_{\mathcal{S}}(t),
\label{Eq:CCE}
\end{equation}
where the product is over all clusters.
Applying this definition to $L_{\mathcal{S}}(t)$ and factoring out
$\tilde{L}_{\mathcal{S}}(t)$, an explicit form for the
$\tilde{L}_{\mathcal{S}}(t)$ is obtained in terms of the
${L}_{\mathcal{S}}(t)$ and the $\tilde{L}_{\mathcal{C}}(t)$
in subsets $\mathcal{C}$ of $\mathcal{S}$,
\begin{equation}
\tilde{L}_{\mathcal{S}}(t) = L_{\mathcal{S}}(t) / \prod_{\mathcal{C} \subset \mathcal{S}} \tilde{L}_{\mathcal{C}}(t).
\label{Eq:ClusterTermExplicit}
\end{equation}
The problem of calculating $L(t)$ is reduced into independent components each
for a distinct cluster of bath spins. The exact solution to $L(t)$ is obtained
if the $\tilde{L}_{\mathcal{S}}(t)$ from all clusters are combined using
Eq.~(\ref{Eq:CCE}) and the approximation to $L\left(t\right)$ up to a maximum
cluster size of $k$ is defined as
\begin{equation}
L^{(k)}(t) = \prod_{|\mathcal{S}| \leq k} \tilde{L}_{\mathcal{S}}(t),
\label{Eq:CCEApprox}
\end{equation}
which involves calculating reduced problems for all clusters each containing
at most $k$ spins. The $L^{(k=2)}(t)$ (2-cluster) calculation is a good
approximation to $L(t)$ when considering only dipolar interactions in the bath
affecting the spin echo, as these are at most a few kHz and hence
perturbative compared to the SHF interactions in the MHz range involving the
donor electron. The CCE is exact, but not always convergent.
We calculated the 2-cluster ($k=2$) approximation to
the CCE and obtained convergence for up to $k=3$.

\begin{figure}[h!]
        \includegraphics[width=3.375in]{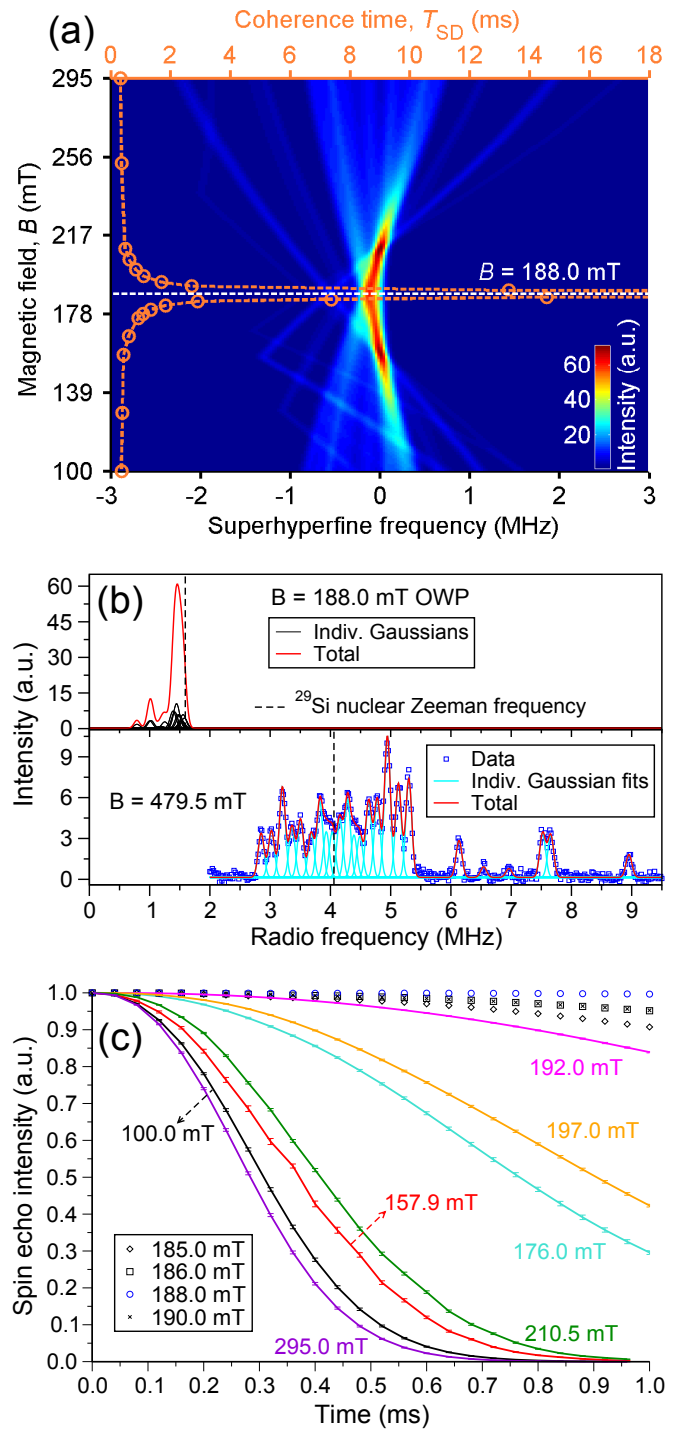}
    \caption{
    (Color online) Suppression of Bi-$^{29}\text{Si}$ spin bath decoherence
    for the $|12\rangle \to |9\rangle$ EPR transition.
    (a) Simulated ENDOR and nuclear spin diffusion
    coherence times $T_{\text{SD}}$ as a function of magnetic field $B$,
    showing collapse of the superhyperfine couplings and a sharp increase
    in $T_{\text{SD}}$ as the field approaches the $B=188.0$~mT optimal working
    point (OWP). The dashed line is a fit.
    (b) Simulated ENDOR at the $B=188.0$~mT OWP (upper panel)
    and experimental spectrum at 9.755~GHz (lower panel).
    (c) Calculated donor Hahn spin echo decays from which coherence
    times in Fig.~\ref{Fig:Suppression}(a) were extracted.
    }
    \label{Fig:Suppression}
\end{figure}

The Hahn echo sequence evolves the combined system-bath state to time $t=2\tau$:
\begin{equation}
|\psi (t=2\tau) \rangle = e^{-i\hat{H}\tau} \left(\hat{\sigma}_{x} \otimes \mathds{1}_{\text{B}}\right)  e^{-i\hat{H}\tau} |\psi(t=0)\rangle,
\label{Eq:HahnEcho}
\end{equation}
where $\hat{\sigma}_{x}$ is the Pauli-$X$ gate acting on the donor and
$\mathds{1}_\text{B}$ denotes the bath identity.
We assume that the time taken for a $\pi$-pulse is small compared to $\tau$.
The initial state was written as a product of the initial donor and bath
states, the former chosen as an equal superposition of states $|12\rangle$ and
$|9\rangle$. The donor subsystem is recovered after tracing over the bath and
the modulus of the normalized off-diagonal element of the donor reduced density
matrix is proportional to the intensity of the echo at time $t=2\tau$. 
The reduced problem of the 2-cluster bath was solved for each of the four
initial 2-product bath states and the average intensity obtained. 2-clusters
were formed by pairing $^{29}\text{Si}$ spins separated by up to the 3rd
nearest neighbor distance in a diamond cubic lattice of side length 160~\AA.
Convergence was obtained as the lattice size and the separation between the
two bath nuclei were extended. It was assumed that $B$ was large enough to
conserve the total $^{29}\text{Si}$ Zeeman energy. Thus, the dipolar
interaction between the two bath spins took the form of a combination
of Ising ($\hat{I}^{z}_{1} \hat{I}^{z}_{2}$) and flip-flop
($\hat{I}_{1}^{+} \hat{I}_{2}^{-} +  \hat{I}_{1}^{-} \hat{I}_{2}^{+}$) terms.
The Kohn-Luttinger electronic wavefunction was used to calculate the isotropic
Fermi contact SHF strength with an ionization energy of 0.069~eV for the
bismuth donor. Calculated couplings were of the same order as those obtained
from data. The data suggest that isotropic couplings predominate; hence
anisotropic couplings were neglected and the simulations were largely insensitive
to orientation. Finally, we obtained the average $L^{(k=2)}(t)$ over 100
spatial configurations of $^{29}\text{Si}$ occupying 4.67\% of lattice sites.
The resulting decay curves were fitted to
$\text{exp}[-t/T_{2} - (t/T_{\text{SD}})^n ]$,
obtaining $T_{2}\gg T_{\text{SD}}$ and values of $n \simeq 2 - 3$.

\section{Suppression of nuclear spin diffusion}\label{Sec:Suppression}
The results of our CCE simulations are presented in 
Fig.~\ref{Fig:Suppression}. Figure~\ref{Fig:Suppression}(a) shows the behavior
around the $B=188.0$~mT OWP, associated with the 
$|12\rangle \to |9\rangle$ EPR line.
The calculated coherence time $T_{\text{SD}}$ (orange dashed line)
is superposed on a color map showing the SHF spectrum:
The latter shows ENDOR spectra simulated as a function of $B$, using
Eq.~(\ref{Eq:IsoENDOR}) and centered about the $^{29}\text{Si}$ nuclear Zeeman
frequency. Strikingly, as $B$ approaches the OWP, the ``comb''
of SHF lines narrows to little more than the width of a single line.
This suggests a drastic reduction in the value of the SHF couplings,
indicating that the bismuth has become largely decoupled from the
$^{29}\text{Si}$ spin bath.

The collapse in the SHF couplings is illustrated further in
Fig.~\ref{Fig:Suppression}(b). The lower panel shows
the measured spectrum at 9.755~GHz. Using our experimentally determined SHF couplings,
the corresponding spectrum at the OWP is shown in the upper panel
of Fig.~\ref{Fig:Suppression}(b), demonstrating clearly the narrowing of the spectrum
[corresponding to the same parameters as Fig.~\ref{Fig:Suppression}(a) but at
the precise field value of the OWP].

The behavior of $T_{\text{SD}}$ is also quite striking and unexpected:
The coherence time predicted by CCE simulations increases asymptotically
at the OWP by several orders of magnitude. Away from the OWP, the results
agree well with experimentally measured values of approximately 0.7~ms.
\cite{George2010}
In Ref.~\onlinecite{George2010}, in a regime of weak state-mixing, simulations
using an effective gyromagnetic ratio indicated that $T_{\text{SD}}$ was
slightly reduced (by about 5\%) in a regime corresponding to lower $df/dB$.
The present study, on the other hand (which in contrast to
Ref.~\onlinecite{George2010} employed a full treatment of the quantum
eigenstate mixing), shows rather an effect very sharply peaked about the OWP:
Nuclear spin diffusion is predicted to be largely suppressed, but over an extremely 
narrow magnetic field range.

Figure~\ref{Fig:Suppression}(c) shows a sample
of CCE spin echo decays from which $T_{\text{SD}}$ times were extracted,
and also serves to further illustrate the sharp increase in $T_{\text{SD}}$.
Similar suppression is present for other OWPs in Si:Bi.
There are $df/dB=0$ minima for the $|15\rangle \to |6\rangle$,
$|14\rangle \to |7\rangle$, $|13\rangle \to |8\rangle$,
$|12\rangle \to |9\rangle$, and $|11\rangle \to |8\rangle$
transitions in the frequency range $5-7.5$~GHz and two maxima for
$|12\rangle \to |11\rangle$ and $|9\rangle \to |8\rangle$ close to $1$~GHz.
The decoupling from the spin bath is also expected to lead to suppression of
decoherence arising from the interaction with a bath of donors.
\cite{Mohammady2010}

\section{Conclusions}\label{Sec:Conclusion}
In conclusion, we present measurements of the SHF couplings between a bismuth
donor and a bath of $^{29}\text{Si}$ impurities which suggest that isotropic couplings dominate.
We further demonstrate the suppression of couplings at OWPs.
Finally, the spin echo decay of the donor is calculated as a many-body problem
and sharp divergence of the spin diffusion time is found at an OWP.
Our study motivates experimental EPR studies in the range $5 - 7.5$~GHz
corresponding to the regions of suppressed decoherence.

~

\section*{ACKNOWLEDGMENTS}
We acknowledge Bernard Pajot for the Si:Bi crystal used here and
thank the EPSRC for financial support through the COMPASSS grant.
M.B.A.K. is supported by a Wellcome Trust studentship.
G.W.M is supported by the Royal Commission for the Exhibition of 1851.
Sandia National Laboratories is a multiprogram laboratory operated by
Sandia Corporation, a wholly owned subsidiary of Lockheed Martin Corporation,
for the U.S. Department of Energy's National Nuclear Security Administration
under Contract No. DE-AC04-94AL85000.


\begin{thebibliography}{99}
\bibitem{Divergence} In the sense that decay of the spin echo is no longer discernible in our converged cluster calculations.
\bibitem{Davies} Although Mims ENDOR offers higher selectivity to small SHF couplings, it would suffer from blind spots.
\end{thebibliography}
\end{document}